**A wrong ground-state structure of $HfO_2$ predicted by machine-learning interatomic potentials based on the PBE functional**

*Shuqi Tang[†], Jinchen Wei[†], Kang Wang, Junjie Zhou, Yihan Zhang, Menglin Huang*, Shiyou Chen**

[†]These authors contributed to this work equally.
S. Tang, J. Wei, K. Wang, J. Zhou, Y. Zhang, M. Huang, S. Chen
College of Integrated Circuits and Micro-Nano Electronics, and Key Laboratory of Computational Physical Sciences (MOE), Fudan University, Shanghai 200433, China
E-mail: menglinhuang@fudan.edu.cn; chensy@fudan.edu.cn

**Abstract**

Machine-learning interatomic potentials (MLIPs) have become powerful tools for material simulations. Many MLIPs are trained based on density functional theory (DFT) datasets generated with the Perdew–Burke–Ernzerhof (PBE) exchange-correlation functional. Using a PBE-based MLIP for $HfO_2$, we identify a previously unreported low-energy $I4_1/amd$ structure, which is predicted to be more stable than the well-known ground-state structure, the monoclinic $P2_1/c$ structure. Since experiments show clearly that $HfO_2$ takes the $P2_1/c$ structure as the ground state, this is obviously a wrong prediction. Unfortunately, the same prediction is also made by widely used PBE-based foundation models such as NequIP-OAM-L and MatterSim-v1-5M. Comparisons among different DFT functionals show that this error originates from the PBE functional, which overstabilizes low-density structures containing sixfold Hf–O octahedral units, such as the $I4_1/amd$ and *Pbcn* phases. The error also affects the calculated energy landscapes and barrier heights along ferroelectric $HfO_2$ polarization switching paths when there are large lattice relaxations. Fortunately, the error can be largely suppressed by other functionals such as PBEsol and local density approximation. Our study serves as a warning about the impact of errors in exchange-correlation functional approximations on the reliability of MLIP simulations of crystal structures and phase transitions.

## Introduction

Machine-learning interatomic potentials (MLIPs) have become an essential tool in materials modelling, combining first-principles accuracy and the computational efficiency required for large-scale atomistic simulations. Neural-network-based MLIPs—encompassing specialized potentials based on advanced architectures (such as SchNet[1,2], DeepMD[3], and Allegro[4]) and pre-trained foundation models (such as M3GNet[5], MACE-MP[6,7], MatterSim[8] and NequIP[9])—have demonstrated a remarkable ability to reproduce reference DFT energies and atomic forces while substantially reducing computational cost, thereby enabling large-scale simulations with near–first-principles accuracy. This capability has driven their widespread deployment across diverse material systems[10-16]. For example, MLIPs have been applied to a wide range of $HfO_2$-related studies, including the amorphous structures[17], phase transitions analysis[18], oxygen-vacancy formation-energy distributions in a-$HfO_2$[19] and radiation-induced displacement damage[20]. These MLIP-driven studies have extended the spatial and temporal scales of simulations and have significantly advanced our understanding of $HfO_2$, which is a key material for advanced microelectronic devices.

The predictive performance of MLIPs critically depends on the quality of the training datasets, which are typically generated from density functional theory (DFT) calculations and are therefore inherently influenced by the choice of the approximation to the exchange–correlation functional. One of the most widely adopted approximations in contemporary DFT calculations[21] and MLIP training dataset generation[6,8] is the Perdew–Burke–Ernzerhof (PBE) generalized gradient approximation (GGA), owing to its favorable computational efficiency and numerical stability[6,8,9]. Notably, the MLIP studies[17-20] on $HfO_2$ mentioned above also employed the PBE functional to generate the training datasets. However, given the strong polymorphic competition in $HfO_2$ (e.g., the monoclinic structures with $P2_1/c$ and $P2_1/m$ symmetries (**Figure 1**(b) and 1(c)), the orthorhombic structures with $Pca2_1$, *Pbcn*, and *Pbca* symmetries (Figure 1(d)–1(f)), the tetragonal structure with $P4_2/nmc$ symmetry (Figure 1(g))[22-24]) and its high sensitivity to external fields and strain[24], a natural question arises: can PBE and PBE-trained MLIPs reliably describe the small energy differences among competing phases and the intricate energy landscapes of $HfO_2$ across diverse physical scenarios?

In this work, we systematically investigate the reliability of PBE-based MLIPs in describing the relative energetics of various $HfO_2$ phases. By performing global structure search, we reveal a serious misprediction, i.e., PBE-based models incorrectly predict an $I4_1/amd$ phase as the ground state, rather than the experimentally established monoclinic $P2_1/c$ ground-state. This misprediction is consistently made by both our specifically trained MLIP and several widely used PBE-based models (e.g., NequIP-OAM-L[9,25] and MatterSim-v1-5M[8]). Through a comparative analysis across different exchange-correlation functional approximations, we trace the origin of this error to systematic errors in the PBE energy landscape. In particular, PBE overstabilizes low-density structures with sixfold Hf–O octahedral units. Furthermore, we demonstrate that these biases affect the calculated energy landscapes and polarization switching barriers in ferroelectric $HfO_2$ under large lattice relaxation. Fortunately, these functional-driven errors are not inevitable. Alternative functionals, such as PBEsol and LDA, can effectively resolve this issue. Overall, our study gives a warning about the impact of errors in exchange-correlation functional approximations on the reliability of MLIP simulations of crystal structures and phase transitions.

## Results

### Identification of an Incorrect PBE Ground State of $HfO_2$

In the past decade, a series of studies have employed PBE-based MLIPs to investigate the structural stability and phase transition behaviors of $HfO_2$[23]. To evaluate whether these MLIPs can accurately describe the energies across different $HfO_2$ structures correctly, we trained an MLIP based on the PBE dataset calculated using the Vienna Ab initio Simulation Package (VASP) code, and conducted a global structural search using the CALYPSO software. In principle, if the identified lowest-energy structure agrees with the well-known ground-state structure with $P2_1/c$ symmetry[26,27] in experiments, the PBE-based MLIPs can be regarded as demonstrating the baseline predictive capability to describe the ground-state structure correctly. Surprisingly, our search identified a low-energy structure with $I4_1/amd$ symmetry, as shown in Figure 1(j). This structure exhibits lattice parameters of approximately a = b = 4.01 Å and c = 10.59 Å, with a unit cell volume of 42.5 $Å^3$/f.u., and its atomic arrangement closely resembles that of rutile $TiO_2$[28] (**Figure S1**). As far as we know, such a low-energy *$I4_1/amd$* structure has not been previously reported for $HfO_2$. Notably, the MLIP calculation shows that its energy is about 17 meV/f.u. lower than that of the monoclinic structure with $P2_1/c$ symmetry, indicating it is at the global minimum on the energy landscape described by the PBE-based MLIP. However, given that experimental evidence definitively confirmed the *$P2_1/c$* structure as the ground state of $HfO_2$[26,27], this result represents an incorrect prediction of the calculation. Such a discrepancy challenges the reliability of the PBE-based MLIP in correctly describing the energies of different $HfO_2$ phases.

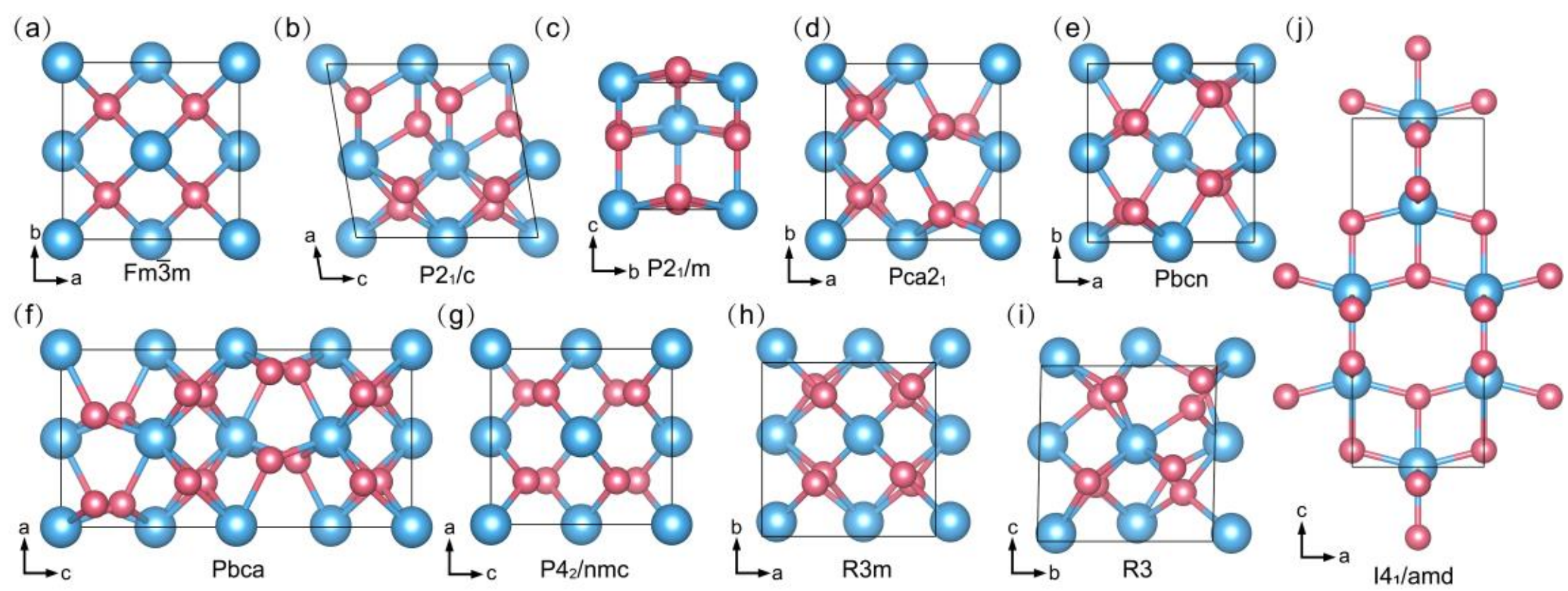

**Figure 1.** Schematic representations of the *$I4_1/amd$* structure and several $HfO_2$ structures. Blue and red spheres represent Hf and O atoms, respectively.

To rule out possible errors arising from our specific MLIP training, we performed additional structural searches using the PBE-based MLIPs publicly available (e.g., NequIP-OAM-L[9,25], MatterSim-v1-5M[8], and MACE-MP-0[6,7,29]). Remarkably, the *$I4_1/amd$* phase remained the most stable structure across these diverse models. These results indicate that the emergence of this "new ground state" is not unique to our MLIP, but represents a general feature observable across multiple PBE-based MLIPs. Consequently, a critical question arises: does the error originate from the machine learning fitting of the interatomic potentials, or is it inherited directly from the training dataset? To address this, we calculated the relative energies of various $HfO_2$ crystal structures using both DFT (PBE) and MLIPs (**Figure 2a-c**). The comparison reveals that, despite minor discrepancies in the energy rankings of certain higher-energy phases, the MLIPs faithfully reproduce the direct DFT (PBE) results, which also predict the $I4_1/amd$ phase as the global energy minimum. The results definitively prove that the wrong prediction of the lowest-energy $I4_1/amd$ phase is not a machine-learning fitting error, but rather an intrinsic error rooted in the PBE functional itself.

Since the wrong ground state originates from the DFT training data, we further investigated whether it represents a general limitation of DFT or a specific misprediction of the PBE functional. As Figure 2d-e shows, we calculated the relative energies of various $HfO_2$ crystal structures using different exchange-correlation functionals, including PBE, PBE-vdW, SCAN, PBEsol, and LDA. Notably, the $I4_1/amd$ phase emerges as the lowest-energy structure exclusively under the PBE functional, confirming that this wrong ground state is unique to the PBE energy landscape. Furthermore, the relative stabilities of the $I4_1/amd$ and *Pbcn* structures exhibit extreme sensitivity to the functional choice. Within the PBE framework, the $I4_1/amd$ phase lies 17 meV/f.u. below the $P2_1/c$ phase. However, its relative energy progressively increases across PBE-vdW, SCAN, PBEsol, and LDA, ultimately becoming the highest-energy structure among all considered phases in the LDA calculation. A similar trend of energy variation is observed for the *Pbcn* structure: it ranks as the third-lowest energy phase in PBE functional (0.05 eV/f.u. higher than the $P2_1/c$ phase) but rises significantly to the second-highest under the LDA functional (0.18 eV/f.u. higher than the $P2_1/c$ phase).

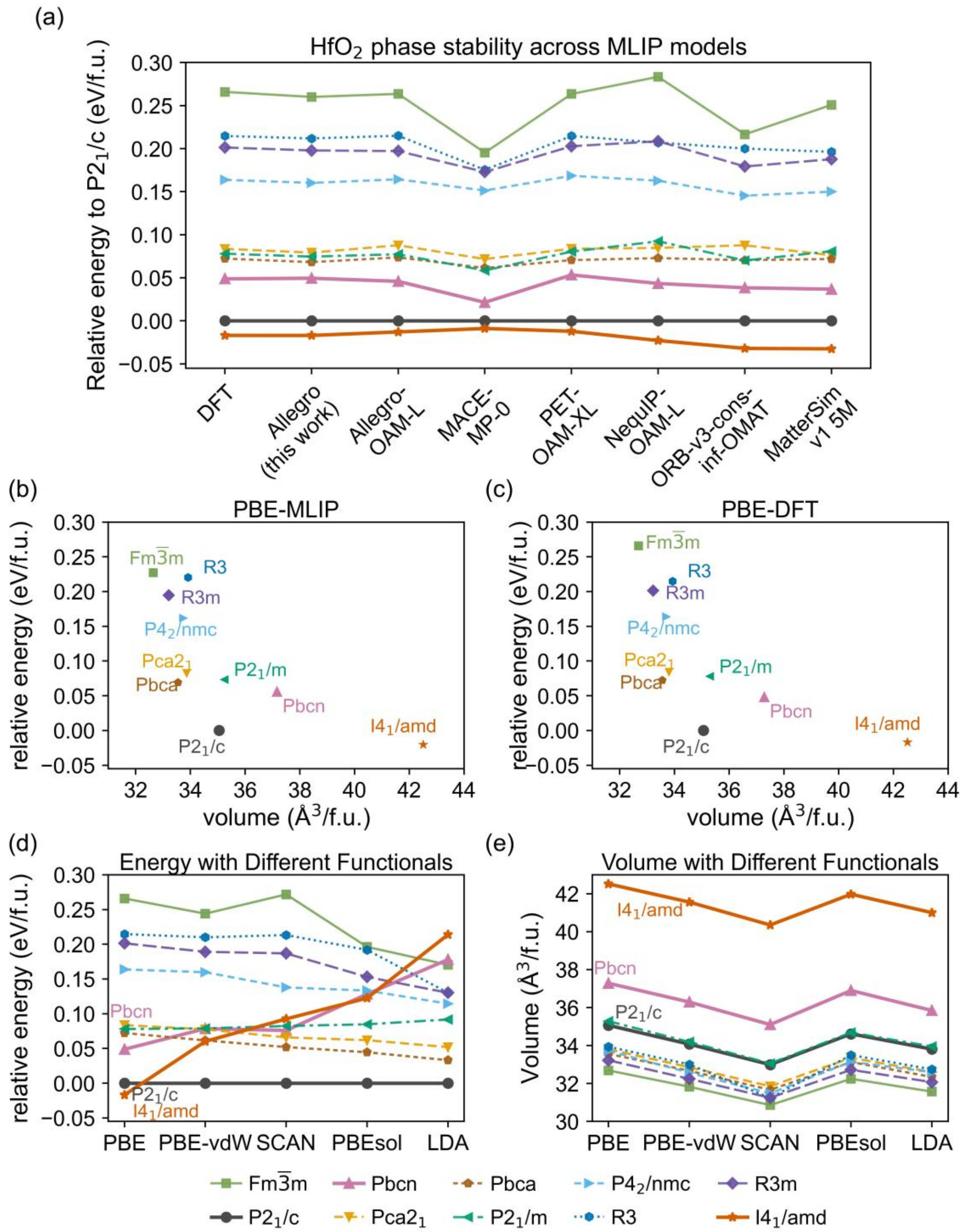

**Figure 2.** Energy ranking and functional dependence of $HfO_2$ crystal structures predicted by PBE-DFT and PBE-based MLIPs. (a) Energy ranking of $HfO_2$ crystal structures obtained from first-principles DFT calculations using the PBE functional and several public PBE-based MLIPs. DFT denotes the PBE first-principles reference calculations; Allegro (this work) refers to the $HfO_2$-specific PBE-based MLIP developed in this study; Allegro-OAM-L[25], MACE-MP-0[6,7,29], PET-OAM-XL[30], NequIP-OAM-L[9,25],ORB-v3-cons-inf-OMAT[31], and MatterSim v1 5M[8] are representative public foundation MLIPs selected from recently developed models benchmarked on Matbench Discovery and released through their corresponding model

repositories. (b) Relative energies and volumes of the $I4_1/amd$ structure and representative $HfO_2$ phases predicted by the PBE-MLIP developed in this work. (c) Relative energies and volumes of the same structures calculated using PBE-DFT. (d) Comparison of relative energies obtained using different exchange–correlation functionals. (e) Comparison of phase volumes obtained using different exchange–correlation functionals. In panels (b–d), the monoclinic m phase is used as the energy reference.

Furthermore, we analyzed the structural characteristics of the *$I4_1/amd$* and *Pbcn* phases to better understand the origin of their functional sensitivity. Although an analysis of bond lengths and angles reveals no clear common structural features (**Figure S2**), both of these phases exhibit larger volumes than other structures considered (Figure 2e) and distinct coordination environments: Hf atoms are six-coordinated and O atoms are three-coordinated, forming local Hf–O octahedral units (**Figure 3**). The relatively large equilibrium volumes and the associated low-coordination environments are likely responsible for the strong functional sensitivity observed in these structures. Equilibrium volumes show a clear systematic dependence on the choice of exchange–correlation functional: PBE favors the largest equilibrium volumes, whereas LDA yields the smallest. As a result, relatively low-density and low-coordination structures such as *$I4_1/amd$* and *Pbcn* are energetically stabilized at larger volumes under the PBE functional, while functionals favoring more compact structures, such as PBEsol and LDA, penalize these low-density configurations, thereby increasing their relative energies.

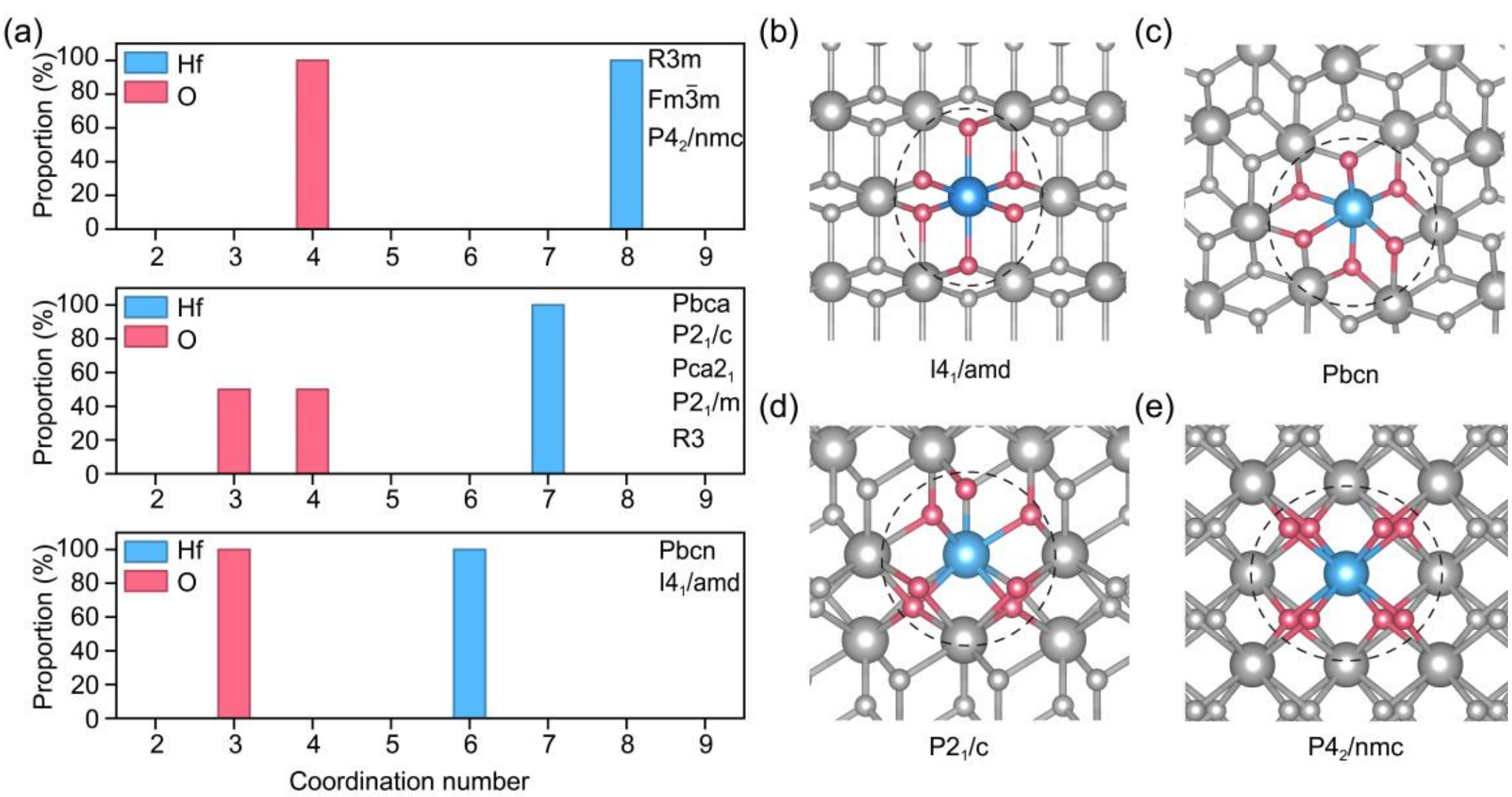

**Figure 3.** Structural characteristics of different $HfO_2$ crystal phases. (a) Distribution of atomic coordination numbers in each phase. (b–e) Polyhedral units centered on Hf atoms in the *$I4_1/amd$*, *Pbcn*, *$P2_1/c$*, and *$P4_2/nmc$* phases, respectively.

Overall, these results indicate that the PBE functional tends to overestimate the stability of large-volume $HfO_2$ phases, thereby spuriously stabilizing a lower-energy symmetric phase on its energy landscape. Such systematic tendencies may be inherited by MLIPs, highlighting the importance of carefully evaluating functional dependence when constructing MLIPs based on DFT calculations using a certain exchange–correlation functional.

**Polarization switching pathways evaluated using different functionals**

To illustrate how the intrinsic biases of the PBE functional affect transition state calculations, we investigate the polarization switching of orthorhombic ferroelectric $HfO_2$ (*$Pca2_1$*). Specifically, we compare four switching paths computed using the PBE and PBEsol functionals, with PBEsol serving as a comparative reference.

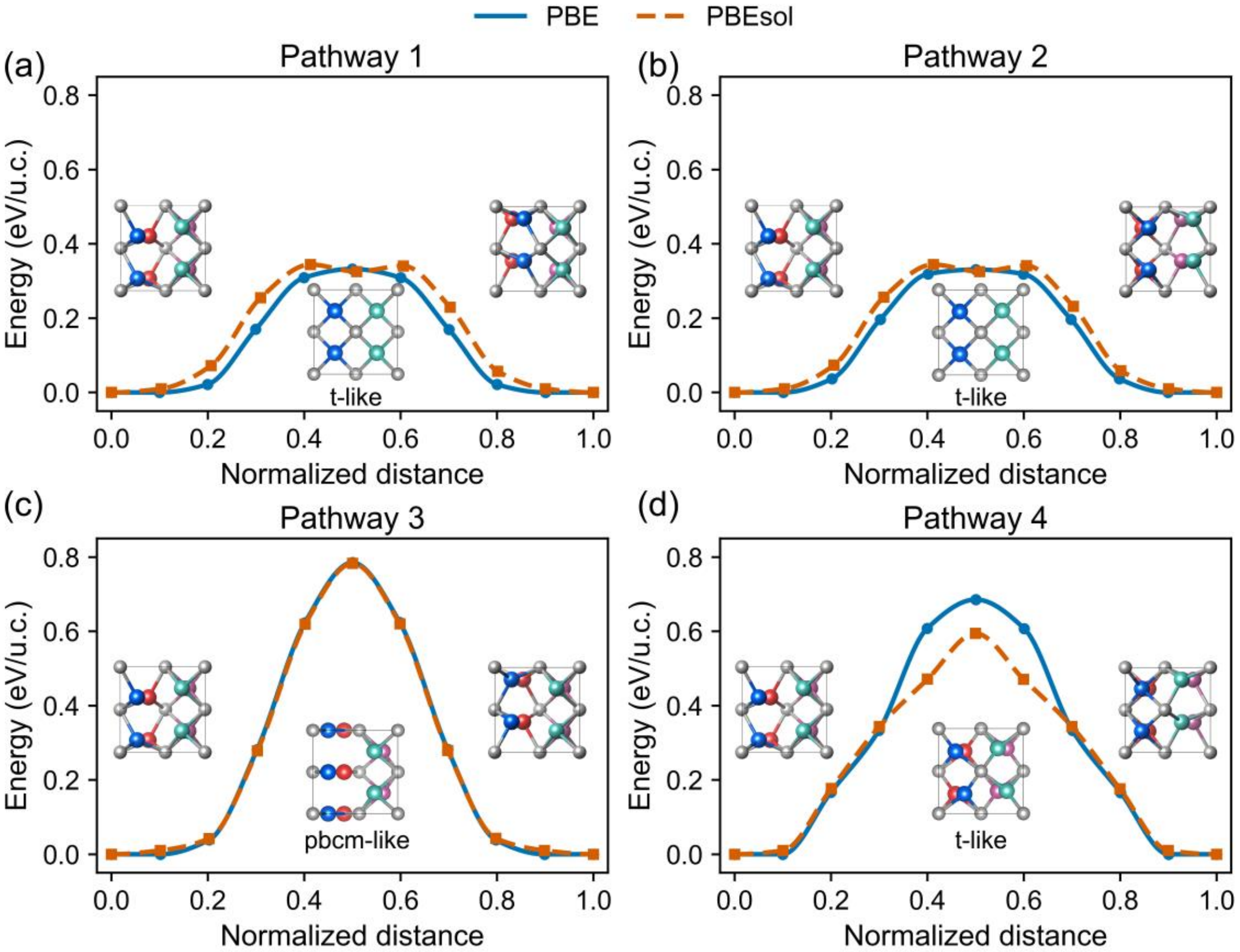

**Figure 4.** Polarization switching energy profiles along four principal pathways under the fixed-lattice condition (with the lattice constants fixed), calculated using PBE and PBEsol functionals. For all four pathways, the transition states predicted by PBE and PBEsol correspond to the same crystal phase.

We first evaluated these four representative pathways [32,33] under the fixed-lattice condition (**Figure 4**). Under this constraint, both functionals yield comparable switching barriers and structurally similar intermediate states. This indicates that when lattice degrees of freedom are artificially suppressed, the overall shape of the potential energy surface is largely consistent between the two functional calculations, enabling both PBE and PBEsol to capture similar key configurations for polarization reversal.

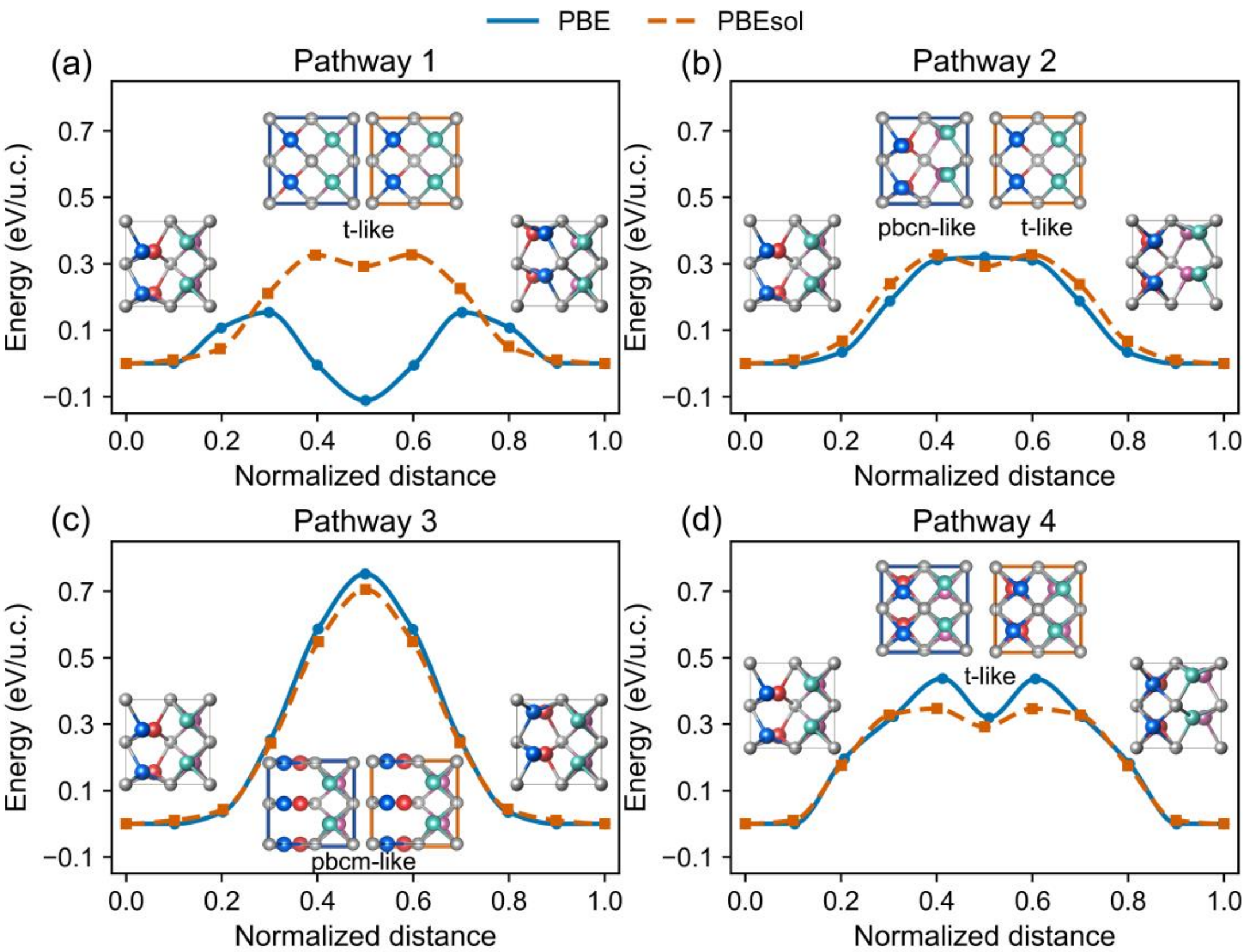

**Figure 5.** Polarization switching energy profiles along four principal pathways under the relaxed-lattice condition, calculated using PBE and PBEsol functionals.

However, once lattice relaxation is allowed, the switching energy profiles exhibit substantial differences in the calculations using the two functionals. As shown in Figure 5, the discrepancies between PBE and PBEsol become substantially larger (especially for pathway2) than those under the fixed-lattice constraint. For pathways 1 and 3, while the predicted intermediate structures remain similar, their corresponding energy profiles exhibit moderate differences. For pathway 4, both functionals predict a tetragonal (t-phase) intermediate, yet their atomic arrangements differ obviously. Pathway 2 shows the most obvious discrepancy: while PBEsol maintains the conventional t-phase transition state, PBE predicts a distinctly different *Pbcn*-like intermediate state.

To explain this behavior, we refer back to the relative energy rankings and volumes of the $HfO_2$ phases (Figure 2). In the PBE functional, the *Pbcn* phase is lower in energy than other intermediate candidates, such as the $P4_2/nmc$ phase. Consequently, when lattice degrees of freedom are released, the system tends to evolve along a lower-energy *Pbcn*-related pathway. In contrast, within the PBEsol framework, the *Pbcn* phase is significantly higher in energy and cannot act as a competitive intermediate state, so the system maintains the *t*-like transition state.

From the perspective of the energy landscape, the system explores a broader configurational space when the volume is taken as a critical degree of freedom. PBE systematically favors larger lattice volumes, whereas PBEsol prefers more compact structural arrangements. As a result, the energy barriers and transition states along polarization switching paths differ between the two functionals. Specifically, within the PBEsol energy landscape, the *Pbcn* phase corresponds to a high-energy region, imposing stronger lattice constraints during polarization switching and resulting in more localized atomic displacements and lattice distortions. By contrast, within the PBE energy landscape, the *Pbcn* phase forms a competing low-energy basin with a lower barrier from $Pca2_1$, making it energetically accessible during relaxed-lattice switching and enabling the emergence of alternative intermediate structures absent under the fixed-lattice condition.

In addition, the choice of functional profoundly affects the local topology of the potential energy surface. Under the fixed-lattice condition, for pathways 1 and 2, PBEsol predicts shallow local minima at the intermediate structures, whereas PBE identifies these corresponding configurations as energy maxima. Similar functional-dependent features have also been reported in previous studies of ferroelectric $HfO_2$[34]. This difference reflects subtle variations in how the two functionals describe local atomic configurations. PBEsol tends to weakly stabilize higher-symmetry intermediates characterized by more uniform interatomic distances. In contrast, PBE yields no such local minima, leaving these symmetric structures stranded at the apex of the kinetic barrier. When lattice relaxation is allowed under the relaxed-lattice condition, the shallow minima in PBEsol become more pronounced for both pathways 1 and 2. This indicates that releasing lattice degrees of freedom amplifies the functional's inherent stabilization of symmetric states.

Our polarization switching calculations demonstrate that while the potential energy surface remains largely consistent under fixed-lattice constraints, structural relaxation with lattice relaxation exposes errors inherent to the PBE functional. Specifically, PBE's systematic overestimation of lattice volumes and its lower energy ranking of the *Pbcn* phase fundamentally change the predicted switching pathways, thereby generating qualitatively different intermediate states and obvious different switching energy profiles. These findings underscore the importance of carefully assessing functional dependence when employing MLIPs trained on DFT-PBE datasets to simulate the processes involving lattice deformation, such as phase transitions or ferroelectric switching dynamics.

**Discussion**

We demonstrated that MLIPs trained on PBE-based DFT datasets incorrectly predict the $I4_1/amd$ structure as the ground-state structure of $HfO_2$, contradicting the experimentally verified monoclinic $P2_1/c$ structure. Comparative DFT calculations reveal that this incorrect prediction—which is also present in widely used foundation models—stems from inherent errors within the PBE exchange-correlation functional. Specifically, the PBE functional underestimates the energy of low-density structures containing sixfold Hf–O octahedral units, such as the $I4_1/amd$ and *Pbcn* phases. Furthermore, we showed that this functional-driven error propagates into the evaluation of ferroelectric polarization switching, significantly altering the calculated energy landscapes and barrier heights when large lattice relaxations are involved.

Our findings provide a cautionary message for MLIP-based materials simulations: an MLIP may faithfully reproduce its reference DFT data while still giving physically unreliable predictions if the underlying exchange–correlation functional is biased. Therefore, the reliability of MLIPs should be evaluated not only by their fitting errors against reference datasets, but also by the physical validity of the exchange–correlation functional used to generate those datasets.

## Methods

### Machine-learning interatomic potential

The MLIP was trained using the Allegro-based implementation described in Ref.[35], which builds on the Allegro equivariant neural network architecture[4]. The training dataset was collected using the on-the-fly machine learning mode[36-38] implemented in the Vienna Ab initio Simulation Package[39] (VASP) during ab initio molecular dynamics (AIMD) simulations. The dataset comprises configurations sampled from multiple $HfO_2$ crystal phases, including the tetragonal (t), monoclinic (m), orthorhombic (o), and Pbcn phases, as well as additional distorted and metastable structures. All AIMD simulations were carried out in the isothermal–isobaric (NPT) ensemble over a temperature range of 100–3300 K. Each trajectory was propagated for 10,000 timesteps to ensure sufficient sampling of diverse local atomic environments and thermally induced lattice distortions. In total, 8,500 configurations were collected for training. The MLIP was trained to reproduce total energies, atomic forces, and stress tensors, with the dataset divided into training and validation subsets at a 9:1 ratio. The trained potential achieves an average energy error of 1.43 meV/atom. All subsequent MLIP-based energy evaluations were carried out using LAMMPS[40] with the Allegro plugin.

### DFT calculation

All DFT calculations were performed using VASP[39]. To systematically assess the influence of exchange–correlation approximations on the structural and energetic properties of $HfO_2$, a representative set of functionals was employed. These include the local density approximation (LDA)[41,42], the generalized gradient approximations (GGAs) in the PBE[21] and PBEsol[43] forms, the meta-GGA functional SCAN[44], as well as the PBE functional augmented with van der Waals interactions (PBE_vdW)[45]. The projector-augmented wave (PAW) method was used to describe the electron–ion interactions, and a plane-wave kinetic energy cutoff of 560 eV was adopted throughout. Brillouin-zone integrations were performed using Γ-centered Monkhorst–Pack k-point meshes, which were chosen according to the cell size and symmetry to ensure convergence; for example, a 6 × 6 × 6 k-point grid was used for 12-atom primitive cells. Electronic self-consistency was achieved with an energy convergence criterion of $1 \times 10^{-6}$ eV, while atomic relaxations were continued until the residual forces on each atom were below 0.01 eV $Å^{-1}$.

**Structural search**

Structural searches were conducted using the CALYPSO[46] package interfaced with the trained MLIP to sample potential local minima across a broad energy landscape. Initial structures were randomly generated with group symmetry constraints while randomizing cell parameters and atomic positions to ensure population diversity. All generated structures were optimized using the PBE-trained MLIP until forces converged to $10^{-3}$ eV/Å, and the lowest-energy configurations were extracted for further analysis.

**Calculation of polarization switching process**

The polarization switching process in ferroelectric $HfO_2$ was calculated using the Nudged Elastic Band (NEB)[47] method. For fixed-lattice paths, the Climbing-Image NEB (CINEB)[48] approach was used to obtain energy–reaction coordinate curves under intrinsic lattice constraints. Fully lattice-relaxed paths were modeled using the Generalized Solid-State NEB (GSSNEB)[49] method to capture possible configurational evolution when lattice relaxation is allowed. Initial and final states correspond to orthorhombic $HfO_2$ structures with opposite polarizations. Complete switching paths were computed using both PBE and PBEsol functionals to systematically compare how different functionals describe the polarization switching energy surfaces and key transition configurations of $HfO_2$.

**Data availability**

The data supporting the findings of this study are included in the main text, Supplementary Information and Source Data files. The Source Data will be made available with the published paper upon acceptance.

## Code availability

The machine-learning interatomic potentials in this work were trained using the Allegro architecture, which is publicly available at https://github.com/mir-group/allegro. The corresponding LAMMPS interface is available at https://github.com/mir-group/pair_nequip_allegro. Custom scripts used for structure processing, energy ranking, switching-path analysis and figure generation are available from the corresponding author upon reasonable request.

## Acknowledgments

This work is supported by National Key Research and Development Program of China (2022YFA1402904), National Natural Science Foundation of China (12334005, 12188101 and 12404089), Science and Technology Commission of Shanghai Municipality (24JD1400600).

## Author contributions

S.T. and J.W. contributed equally to this work. S.T. performed the DFT calculations, MLIP training, MLIP testing, structure search, polarization switching calculations, data analysis and manuscript writing. J.W. contributed to the conceptualization of the study, performed DFT calculations and revised the manuscript. K.W. supervised part of the work and contributed to manuscript writing. J.Z. and Y.Z. contributed to MLIP testing. M.H. and S.C. conceived the project, supervised the research, guided the analysis and interpretation of the results, and revised the manuscript. All authors discussed the results and approved the final version of the manuscript.

**Competing interests**

The authors declare no competing interests.